\documentclass[runningheads]{svmult}

\usepackage{graphicx}  % standard LaTeX graphics tool
\usepackage{physprbb}  % modified textarea for proceedings,
\def\la{\mathrel{\hbox{\rlap{\hbox{\lower4pt\hbox{$\sim$}}}\hbox{$<$}}}}
\def\ga{\mathrel{\hbox{\rlap{\hbox{\lower4pt\hbox{$\sim$}}}\hbox{$>$}}}}

\begin{document}
\title*      {Lens Galaxies vs.\ CDM}
\toctitle    {Lens Galaxies vs.\ CDM}
\titlerunning{Lens Galaxies vs.\ CDM}
\author{Charles R.\ Keeton\inst{1,2,3}}
\authorrunning{Keeton}
\institute{
Steward Observatory, University of Arizona, 933 N.\ Cherry Ave.,
Tucson, AZ 85721
\and
Astronomy and Astrophysics Department, University of Chicago,
5640 S.\ Ellis Ave., Chicago, IL 60637
\and
Hubble Fellow
}

\maketitle              % typesets the title of the contribution

\begin{abstract}
By directly probing mass distributions, gravitational lensing offers
several new tests of the CDM paradigm.  Lens statistics place upper
limits on the dark matter content of elliptical galaxies.  Galaxies
built from CDM mass distributions are too concentrated to satisfy
these limits, so lensing extends the ``concentration problem'' in
CDM to elliptical galaxies.  The central densities of the model
galaxies are too low on $\sim\!10$ pc scales to agree with the
lack of central images in observed lenses.  The flux ratios of
four-image lenses imply a substantial population of dark matter
clumps with a typical mass $\sim\!10^{6}\ M_\odot$.  Thus, lensing
implies the need for a mechanism that reduces dark matter densities
on kiloparsec scales without erasing structure on smaller scales.
\end{abstract}

\section{Introduction}

The popular Cold Dark Matter (CDM) paradigm is facing several
challenges on small scales (e.g., \cite{Moore01}).  The dynamics
of spiral galaxies, especially rotation curves and fast-rotating
bars, suggest that in observed galaxies dark matter halos are much
less concentrated than predicted by CDM (e.g., \cite{Debattista},
\cite{deBlok}), although this conclusion is still controversial
(e.g., \cite{vandenBosch}).  The number of satellite dwarf galaxies
in the Local Group is much smaller than the number of subhalos in
CDM simulations \cite{Klypin}, \cite{Moore99}, although the
discrepancy may be explained by the astrophysics of star formation
rather than by the physics of the dark matter particle
\cite{Bullock00}.  These tests of CDM are limited, however, by
uncertainties in interpreting luminous tracers of the potential.
Gravitational lensing offers a different test that probes mass
distributions directly.  Strong lensing by galaxies robustly
determines the total mass in the inner 5--10 kpc of lens
galaxies, which are predominantly elliptical galaxies.  It also
offers the possibility to detect small-scale mass concentrations
in galaxy halos \cite{Chiba}, \cite{Dalal}, \cite{Keeton2},
\cite{Mao}, \cite{Metcalf}.  Lensing thus offers new tests of CDM
that avoid dynamical uncertainties and extend the tests from spiral
galaxies to ellipticals.

\section{Star+Halo Models}

I construct new models for lens statistics that include both stellar
and dark matter components (see \cite{Keeton} for details).  In
principle, I take a CDM dark matter halo, add baryons, let the
baryons condense into a galaxy, and use the adiabatic contraction
formalism \cite{Blumenthal} to compute how the dark matter
distribution is modified by the baryons.\footnote[2]{Gottbrath
\cite{Gottbrath} shows that adiabatic contraction agrees remarkably
well with detailed gasdynamical simulations even in the merger
scenarios thought to produce elliptical galaxies.} In practice,
I fix the stellar galaxies and use the models to place dark matter
halos around them. The stellar components are treated as Hernquist
models for elliptical galaxies, normalized by observed galaxy
luminosity functions \cite{Lin}, Fundamental Plane relations
\cite{Schade}, and Bruzual \& Charlot \cite{Bruzual} model
mass-to-light ratios (which are reliable for the old stellar
components of elliptical galaxies).

Two free parameters apply to the dark matter halos. First, halos
with the Navarro, Frenk \& White \cite{NFW} dark matter profile are
described by a concentration parameter. A halo's concentration
is determined by its mass and redshift, but with a scatter of
$0.18$ dex \cite{Bullock01}. I include the scatter and take
the median concentration to be a free parameter. Second, to relate
the total, virial mass of the dark matter halo ($M_d$) to the mass
of the stellar component ($M_s$), I define the ``cooled mass
fraction'' $f_{cool}=M_s/(M_d+M_s)$. I take the cooled mass
fraction to be the second free parameter in the models, assuming
only that it is smaller than the global baryon fraction, $f_{cool}
\le \Omega_b/\Omega_M$.

\section{Lens Statistics and Galaxy Masses}

Lens statistics can be used to test the CDM models, because changes
to galaxy dark matter halos affect the number of lenses and the
distribution of lens image separations. Figure~1a demonstrates the
test by comparing the model predictions with the data from the
Cosmic Lens All-Sky Survey (CLASS; e.g., \cite{Helbig}), which is the
largest homogeneous survey for lenses. Increasing the concentration
of dark matter halos raises the amount of dark matter in the inner
parts of galaxies, leading the models to predict more and larger
lenses. Because the stellar components of the galaxies are fixed,
{\it decreasing\/} the cooled mass fraction {\it increases\/} the
amount of dark matter, again leading to more and larger lenses.

\begin{figure}[t]
\centerline{
  \includegraphics[width=6.0cm]{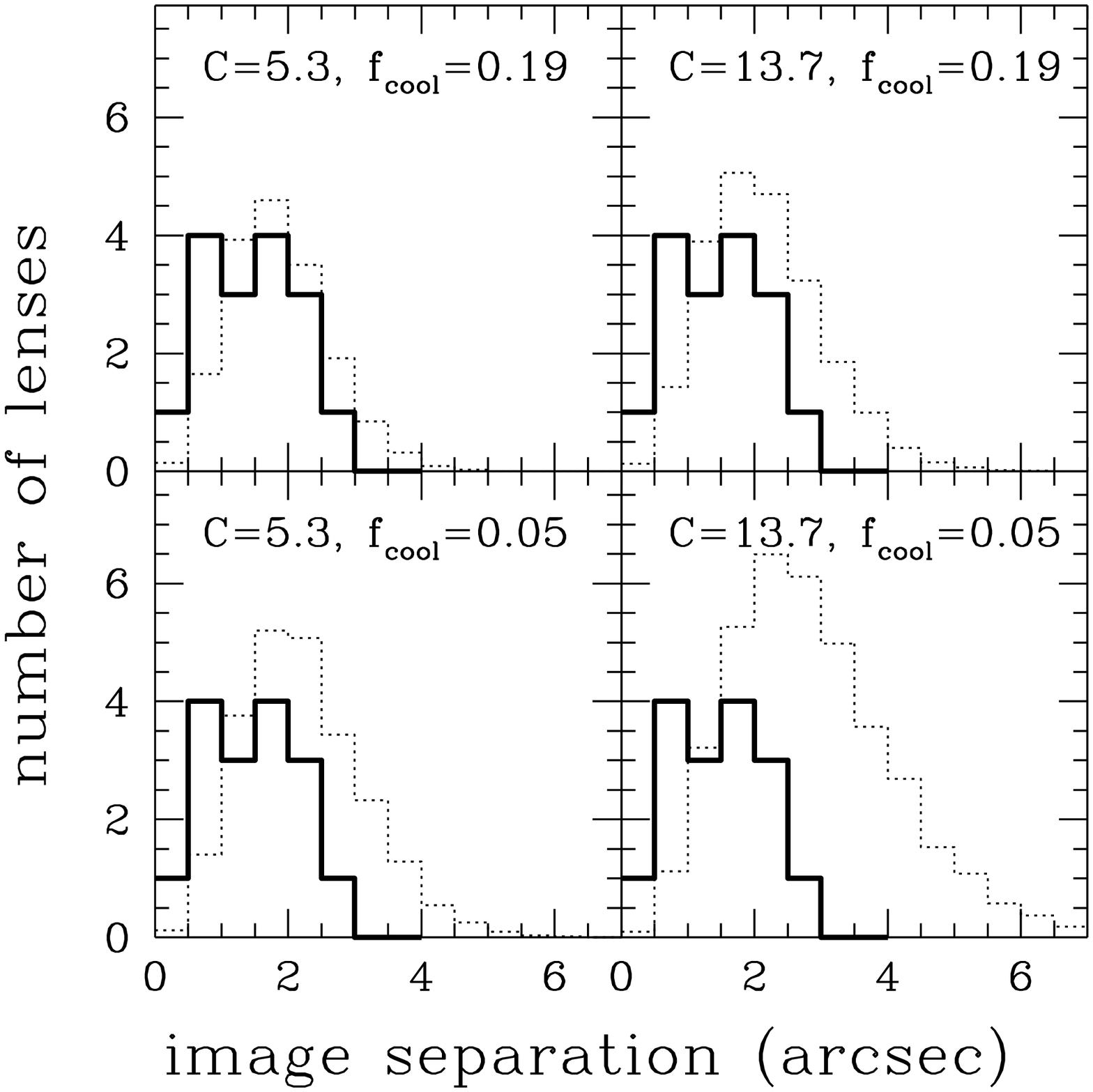}
  \includegraphics[width=6.0cm]{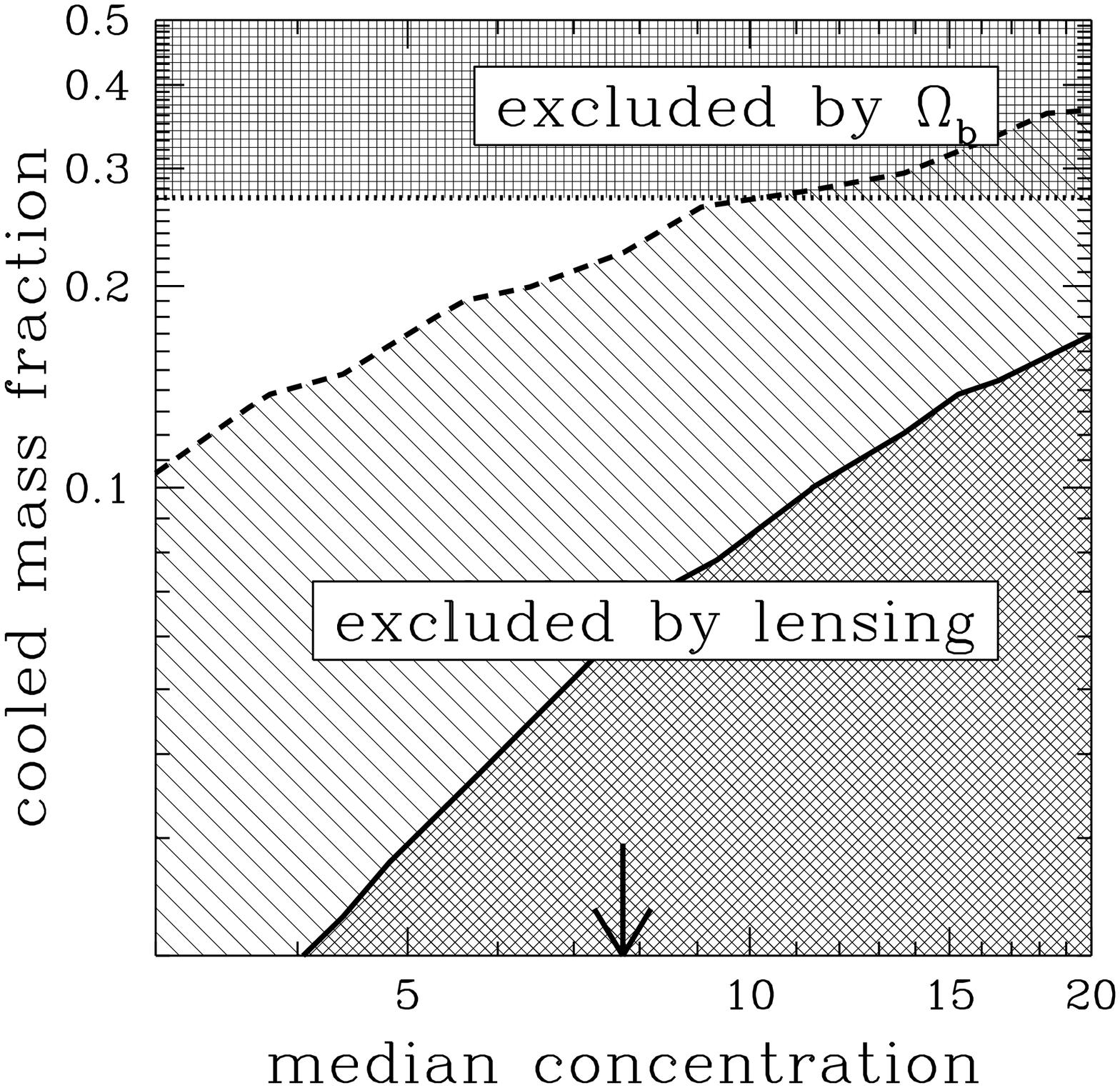}
}
\caption{
{\it (a, left)\/} Image separation histograms for the CLASS data
(solid lines) and for sample models (dotted lines). The model
concentration $C$ and cooled mass fraction $f_{cool}$ are indicated
in each panel.
{\it (b, right)\/} Confidence intervals in the $(C,f_{cool})$
plane. The hatched region is excluded at 95\% confidence by the
distribution of lens image separations, and the cross-hatched
region is further excluded by the number of lenses. The shaded
region at the top is excluded at 95\% confidence by measurements of
$\Omega_b$ (e.g., \cite{deBernardis}, \cite{Tytler}).
All results are shown for a cosmology with $\Omega_M=0.2$ and
$\Omega_\Lambda=0.8$. See \cite{Keeton} for more discussion.
}
\end{figure}

Using statistical tests to compare the models to the data leads
to confidence intervals in the $(C,f_{cool})$ plane, as shown in
Fig.~1b. Lensing requires the models to have low concentrations or
high cooled mass fractions. Adding the constraint on $f_{cool}$
from the baryon content of the universe leaves only a small region
of parameter space where the models are acceptable. Fiducial CDM
models predict a median concentration $C \simeq 7.7$ for galaxies
(indicated in Fig.~1b). This value is allowed by lens statistics
only if galaxies are nearly 100\% efficient at cooling their
baryons ($f_{cool} \simeq \Omega_b/\Omega_M$), which is implausible
(e.g., \cite{Balogh}). The constraints in Fig.~1b are conservative,
because most of the systematic effects in the lensing analysis
strengthen the lensing constraints (see \cite{Keeton}).  Changing
the cosmology (increasing $\Omega_M$) has little effect on the
lensing analysis but reduces the upper limit
$f_{cool} \le \Omega_b/\Omega_M$.

Translating the constraints into enclosed mass leads to the
conclusion that dark matter can account for no more than 33\% of
the mass within $1\,R_e$ and 40\% of the mass within $2\,R_e$
(95\% confidence limits on average mass fractions).  Note that
these limits are for the mass in {\it spheres\/}, whereas lensing
limits on the mass in {\it cylinders\/} indicate that dark matter
halos are still important in ellipticals.  The lensing limits are
consistent with the mass estimates from dynamical analyses of
nearby elliptical galaxies \cite{Gerhard}. By contrast, the CDM
models predict dark matter mass fractions of $\sim\!28\%$ inside
$1\,R_e$ if baryon cooling is 100\% efficient, and even higher
fractions for more reasonable cooling efficiencies.

\section{Odd Images and Galaxy Centers}

Nearly all observed lenses have an even number of images (usually
two or four). Lens theory, by contrast, predicts that each lens
should have an additional ``odd'' image located near the center of
the lens galaxy, although it is demagnified by high central density
of the lens galaxy. At optical wavelengths an odd image would be
swamped by light from the lens galaxy, but in a radio lens an odd
image should be detectable. The lack of odd images in radio lenses
thus places strong lower limits on the central densities of
lens galaxies \cite{Rusin}.

The CDM model galaxies predict that $\ga\!30\%$ of (radio) lenses
should have detectable odd images, implying that the model
densities are much too low on $\sim\!10$ pc scales (see \cite{Keeton}
for details).  Steep central cusps ($\rho \propto r^{-\alpha}$
with $\alpha \simeq 2$) and/or central black holes can help
suppress odd images, but for realistic parameter ranges neither
offers an attractive solution.  The lack of odd images in observed
lenses thus remains a puzzle whose resolution will reveal
interesting new constraints on the very inner parts of distant
galaxies.

\section{Lensing and CDM Substructure}

One claimed problem with CDM is that the number of subhalos in CDM
model galaxies is much larger than the number of satellite dwarf
galaxies in the Local Group, which suggests that CDM overpredicts
the amount of substructure in galaxy-mass halos \cite{Klypin},
\cite{Moore99}.  Two solutions have been proposed.  On the one
hand, changing the nature of the dark matter could reduce the
power on small scales and eliminate the subtructure \cite{Bode},
\cite{Colin}.  On the other hand, astrophysical processes such
as photoionization could inhibit star formation in low mass systems,
meaning that the CDM subhalos exist but are dark \cite{Bullock00}.
Dwarf galaxy surveys cannot distinguish between these scenarios.
Tidal streams offer an alternate test, because they can be disrupted
by encounters with subhalos \cite{Ibata}, \cite{Mayer}, but the
observational evidence is not yet available.

Lensing offers a better test by being directly sensitive to mass
in subhalos.  Mass clumps in the lens galaxy introduce small-scale
variations in the lensing potential that alter the flux ratios of
the lensed images \cite{Chiba}, \cite{Mao}, \cite{Metcalf}.
Dalal \& Kochanek \cite{Dalal} show that the incidence of
``anomalous'' flux ratios\footnote[3]{Flux ratios that cannot be
explained by smooth lens models.} in 4-image lenses requires that
$\sim\!2\%$ of the mass be in small clumps on the scale
$\sim\!10^{4}$--$10^{8}\ M_\odot$, which is in good agreement with
the amount of substructure predicted by CDM.  In other words, lensing
strongly supports the scenario in which many subhalos exist but lack
stars, and opposes changes to the nature of the dark matter that
eliminates substructure.

To complement statistical analyses like \cite{Dalal}, I have studied
a single 4-image lens in detail using data at a variety of wavelengths
to obtain constraints on individual mass clumps \cite{Keeton2}.  In
B1422+231, the optical A/C flux ratio is largely consistent with
smooth lens models while the radio A/C flux ratio is not (Fig.~2).
Simultaneously explaining the optical and radio flux ratios and the
shape of the radio image requires a mass clump in front of image A.
A highly concentrated, point mass clump must have a mass
$\sim\!10^{4}$--$10^{5}\ M_\odot$, while a more extended isothermal
sphere must have a mass $\sim\!10^{6}$--$10^{7}\ M_\odot$.  This is
the first measurement of a particular clump lying in a distant galaxy
($z_l=0.34$) and detected by its mass.  Interestingly, there also
appears to be a clump passing in front of image B, but this clump is
probably just a star in the lens galaxy.  In the future, detailed
analyses of individual clumps as in B1422+231 will be combined with
statistical analyses like \cite{Dalal} to constrain not only the
substructure mass fraction but also the masses, densities, and
sizes of dark subhalos, and the substructure mass function.

\begin{figure}[t]
\centerline{
  \includegraphics[width=6.0cm]{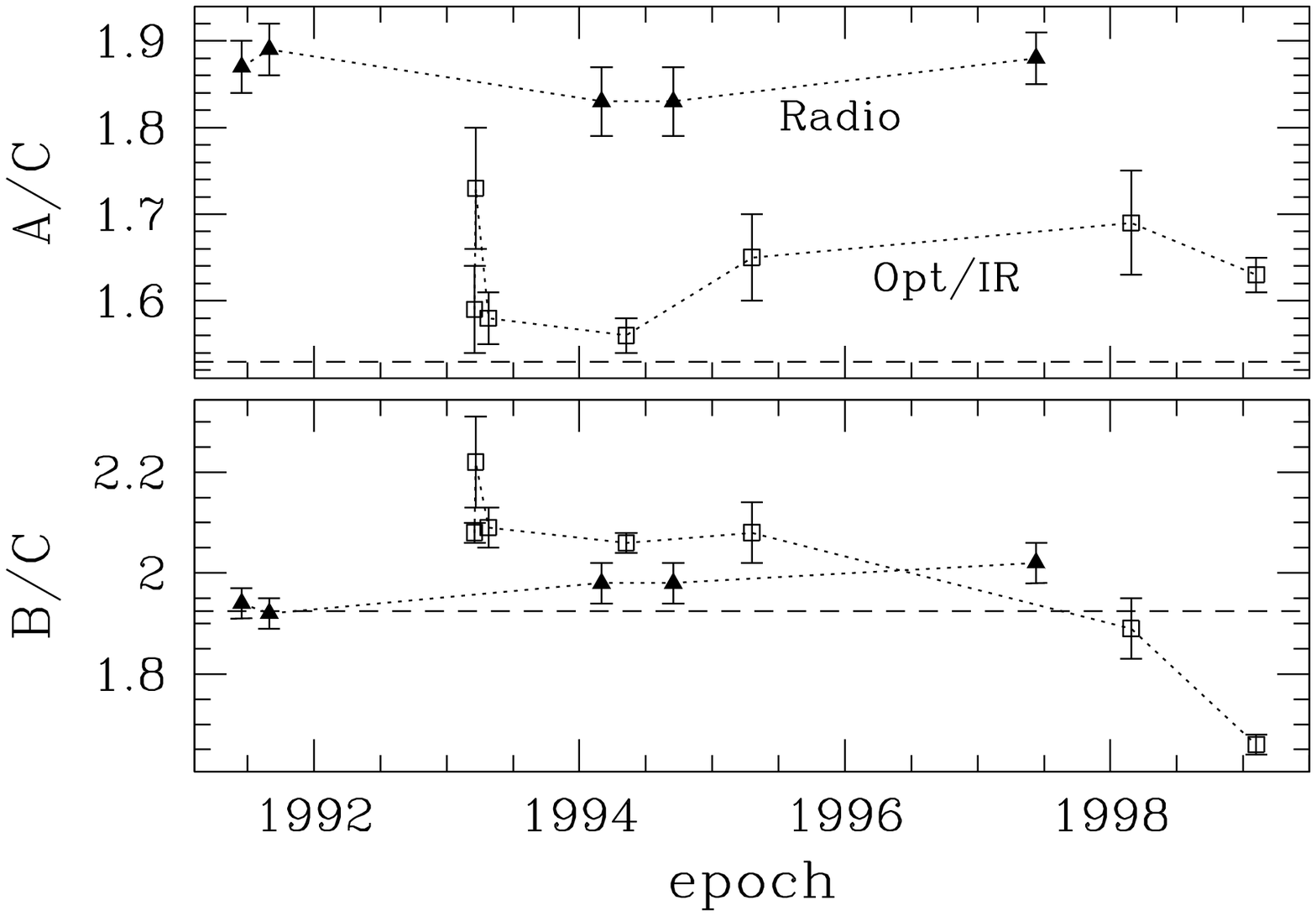}
  \includegraphics[width=6.0cm]{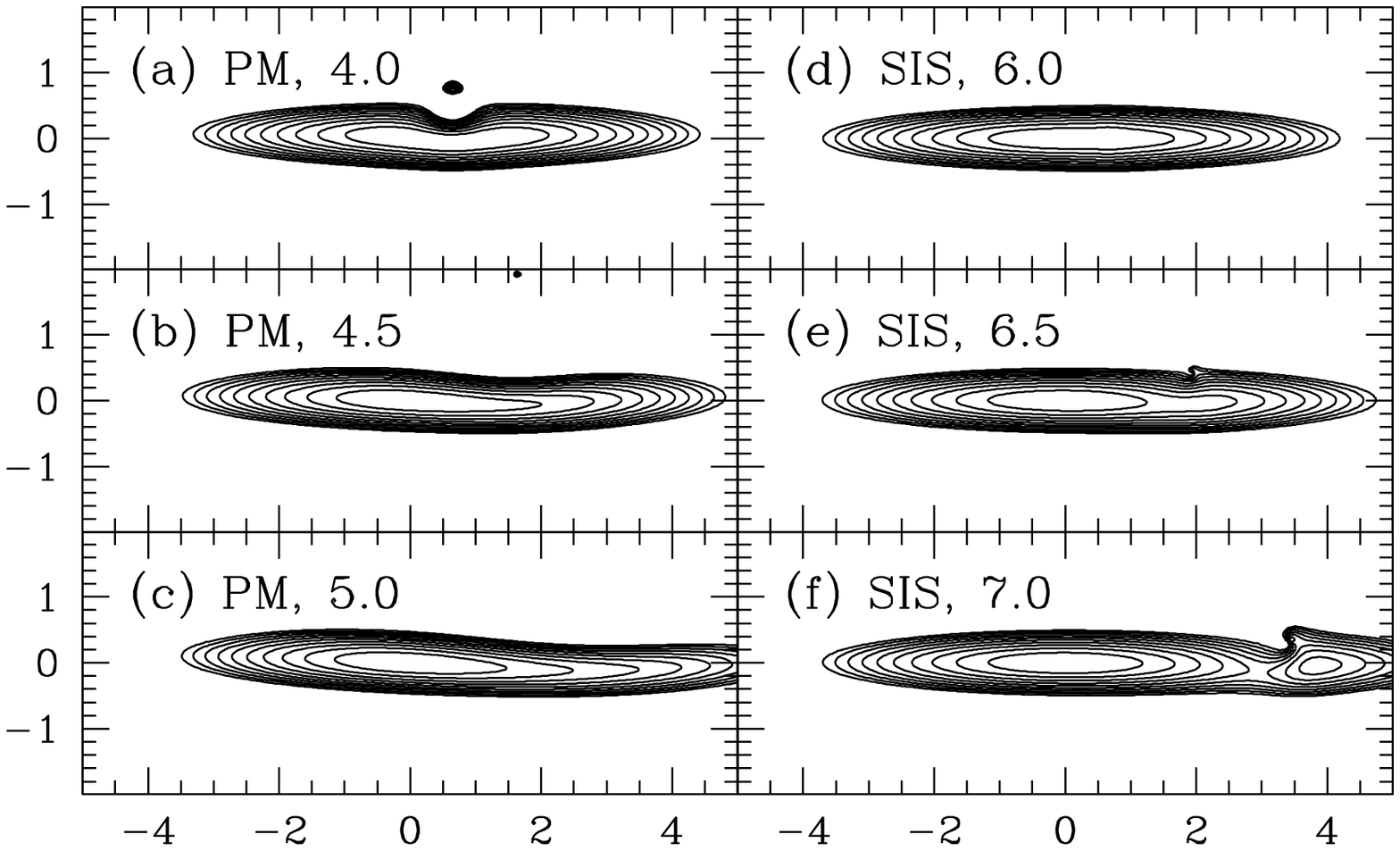}
}
\caption{
{\it (Left)\/} Flux ratios for B1422+231 as a function of time.  The
dashed lines indicate the flux ratios predicted by a smooth lens
model.  See \cite{Keeton2} for details (including references for
the data).
{\it (Right)\/} Maps of the radio image predicted by various
sub-lensing models for image A, assuming infinite resolution.
Results are shown for models with the clump treated as a point mass
(PM) or a singular isothermal sphere (SIS); each panel gives the
clump mass as $\log M$ (in $h^{-1}\,M_\odot$).  The axes are labeled
in mas; the contours are spaced by 0.2 dex.
}
\end{figure}

\section{Conclusions}

Lens statistics imply that the dark matter densities in the inner
parts of elliptical galaxies are lower than predicted by CDM, in
agreement with the conclusion from dynamical analyses of spiral
galaxies. The CDM paradigm must therefore be modified to reduce
dark matter densities on kiloparsec scales. Various mechanisms
have been proposed ranging from astrophysics (disk bars that erase
dark matter cusps \cite{Weinberg}) to cosmology (a tilted power
spectrum \cite{Alam}) to particle physics (dark matter that is
not collisionless and cold \cite{Bode}, \cite{Colin}, \cite{Spergel}).

Lensing also implies that lens galaxies have high densities on
small scales ($\la\!10$ pc). The central densities of galaxies must
be much higher than predicted in CDM model galaxies to explain the
absence of central or ``odd'' images in observed lenses. The flux
ratios in four-images lenses imply that a substantial fraction of
the dark matter ($\sim\!2\%$) lies in small-scale clumps rather
than a smooth halo component \cite{Dalal}, and B1422+231 suggests
that a typical clump mass is $\sim\!10^{6}\ M_\odot$ \cite{Keeton2}.
Thus, while lensing supports other evidence that a mechanism is
needed to reduce dark matter densities on kiloparcsec scales, it
also suggests that the mechanism must {\it not\/} remove structure
on small scales --- which argues against changing the nature of
the dark matter particle.

\vskip\baselineskip
{\bf Acknowledgments.}
Support for this work was provided by Steward Observatory at the
University of Arizona, and by NASA through Hubble Fellowship grant
HST-HF-01141.01-A from the Space Telescope Science Institute, which
is operated by the Association of Universities for Research in
Astronomy, Inc., under NASA contract NAS5-26555.


\begin{thebibliography}{88.}

\bibitem{Alam}
S.~M.~K.~Alam, J.~S.~Bullock, D.~H.~Weinberg: preprint astro-ph/0109392
(2001)

\bibitem{Balogh}
M.~L.~Balogh et al.: MNRAS 326, 1228 (2001)

\bibitem{Blumenthal}
G.~Blumenthal, S.~Faber, R.~Flores, J.~Primack: ApJ 301, 27 (1986)

\bibitem{Bode}
P.~Bode, J.~P.~Ostriker, N.~Turok: ApJ 556, 93 (2001)

\bibitem{Bruzual}
G.~Bruzual, S.~Charlot: ApJ 405, 538 (1993)

\bibitem{Bullock00}
J.~.S.~Bullock, A.~V.~Kravtsov, D.~H.~Weinberg: ApJ 539, 517 (2000)

\bibitem{Bullock01}
J.~.S.~Bullock et al.: MNRAS 321, 559 (2001)

\bibitem{Chiba}
M.~Chiba: preprint astro-ph/0109499 (2001)

\bibitem{Colin}
P.~Colin, V.~Avila-Reese, O.~Valenzuela: ApJ 542, 622 (2000)

\bibitem{Dalal}
N.~Dalal, C.~S.~Kochanek: preprint astro-ph/0111456 (2001)

\bibitem{Debattista}
V.~P.~Debattista, J.~A.~Sellwood: ApJL 493, L5 (1998)

\bibitem{deBernardis}
P.~de~Bernardis et al.: preprint astro-ph/0105296 (2001)

\bibitem{deBlok}
W.~J.~G.~de~Blok et al.: ApJ 552, L23 (2001)

\bibitem{Gerhard}
O.~Gerhard, A.~Kronawitter, R.~P.~Saglia, R.~Bender: AJ 121, 1936 (2001)

\bibitem{Gottbrath}
C.~Gottbrath: MS thesis, University of Arizona (2000)

\bibitem{Helbig}
P.~Helbig: preprint astro-ph/0008197 (2000)

\bibitem{Ibata}
R.~A.~Ibata, G.~F.~Lewis, M.~J.~Irwin: preprint astro-ph/0110690 (2001)

\bibitem{Keeton}
C.~R.~Keeton: ApJ 561, 46 (2001)

\bibitem{Keeton2}
C.~R.~Keeton: preprint astro-ph/0111595 (2001)

\bibitem{Klypin}
A.~Klypin, A.~V.~Kravtsov, O.~Valenzuela, F.~Prada: ApJ 522, 82 (1999)

\bibitem{Lin}
H.~Lin et al.: ApJ 518, 533 (1999)

\bibitem{Mao}
S.~Mao, P.~Schneider: MNRAS 295, 587 (1998)

\bibitem{Mayer}
L.~Mayer et al.: preprint astro-ph/0110386 (2001)

\bibitem{Metcalf}
R.~B.~Metcalf \& P.~Madau: preprint astro-ph/0108224 (2001)

\bibitem{Moore99}
B.~Moore et al.: ApJL 524, L19 (1999)

\bibitem{Moore01}
B.~Moore: preprint astro-ph/0103100 (2001)

\bibitem{NFW}
J.~F.~Navarro, C.~S.~Frenk, S.~D.~M.~White: ApJ 462, 563 (1996)

\bibitem{Rusin}
D.~Rusin, C.-P.~Ma: ApJL 549, L33 (2001)

\bibitem{Schade}
D.~Schade, L.~F.~Barrientos, O.~L\'opez-Cruz: ApJL 477, L17 (1997)

\bibitem{Spergel}
D.~N.~Spergel, P.~J.~Steinhardt: PRL 84, 3760 (2000)

\bibitem{Tytler}
D.~Tytler, J.~M.~O'Meara, N.~Suzuki, D.~Lubin: Phys.Rep.\ 333, 409 (2000)

\bibitem{vandenBosch}
F.~C.~van~den~Bosch, R.~A.~Swaters: MNRAS 325, 1017 (2001)

\bibitem{Weinberg}
M.~D.~Weinberg, N.~Katz: preprint astro-ph/0110632 (2001)

\end{thebibliography}
\end{document}